\documentclass[journal=jacsat,manuscript=article]{achemso}

\usepackage[version=3]{mhchem} 
\usepackage{xcolor}
\usepackage{amsmath}

\author{Wenbo Luo}
\affiliation{Hangzhou Institute for Advanced Study, University of Chinese Academy of Sciences, 310024, Hangzhou, China}
\alsoaffiliation{School of Advanced Interdisciplinary Sciences, University of Chinese Academy of Sciences, No.1, Yanqihu East Rd, Huairou District, 101408, Beijing, China}

\author{Yitong Gu}
\affiliation{Shanghai Institute of Optics and Fine Mechanics, Chinese Academy of Sciences, Shanghai, 201800, China}
\alsoaffiliation{Hangzhou Institute for Advanced Study, University of Chinese Academy of Sciences, 310024, Hangzhou, China}

\author{Jianwei Wang}
\affiliation{Shanghai Institute of Optics and Fine Mechanics, Chinese Academy of Sciences, Shanghai, 201800, China}
\alsoaffiliation{Hangzhou Institute for Advanced Study, University of Chinese Academy of Sciences, 310024, Hangzhou, China}

\author{Fei Yu}
\affiliation{Shanghai Institute of Optics and Fine Mechanics, Chinese Academy of Sciences, Shanghai, 201800, China}
\alsoaffiliation{Hangzhou Institute for Advanced Study, University of Chinese Academy of Sciences, 310024, Hangzhou, China}

\author{Chunlei Yu}
\affiliation{Shanghai Institute of Optics and Fine Mechanics, Chinese Academy of Sciences, Shanghai, 201800, China}
\alsoaffiliation{Hangzhou Institute for Advanced Study, University of Chinese Academy of Sciences, 310024, Hangzhou, China}

\author{Lili Hu}
\affiliation{Shanghai Institute of Optics and Fine Mechanics, Chinese Academy of Sciences, Shanghai, 201800, China}
\alsoaffiliation{Hangzhou Institute for Advanced Study, University of Chinese Academy of Sciences, 310024, Hangzhou, China}

\author{Zhichao Ruan}
\affiliation{School of Physics, State Key Laboratory of Extreme Photonics and Instrumentation, and Zhejiang Province Key Laboratory of Quantum Technology and Device, Zhejiang University, Hangzhou 310027, China}
 
\author{Ning Wang}
\affiliation{Hangzhou Institute for Advanced Study, University of Chinese Academy of Sciences, 310024, Hangzhou, China}
\email{wn@ucas.ac.cn}

\title[An \textsf{achemso} demo]
  {A power-in-bucket model enabled designs of nanostructure-enhanced waveguides for highly efficient wide-angle light couplings}


\begin{document}

\begin{tocentry}

\centering
\includegraphics[scale = 0.31]{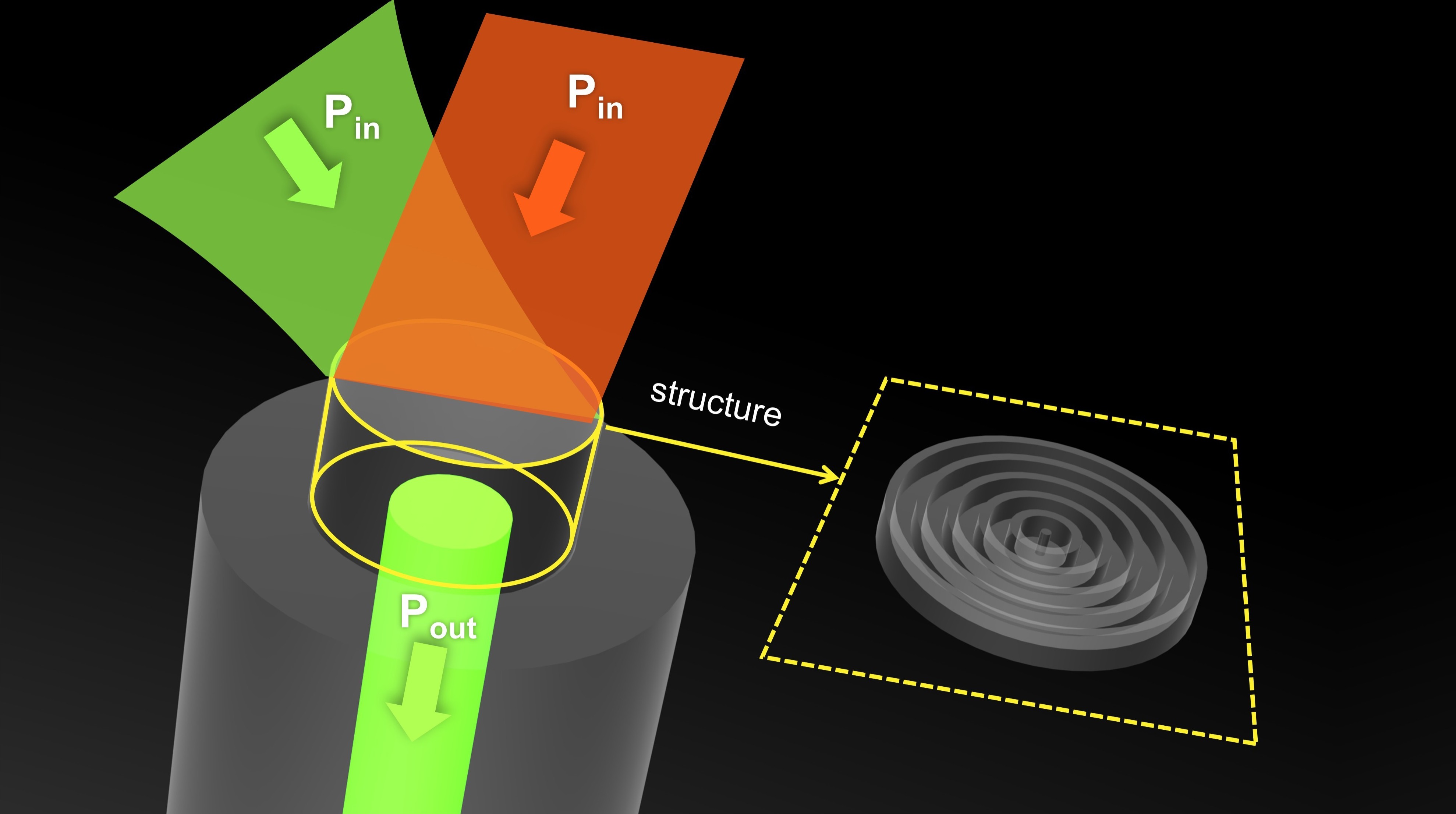}
The diagram illustrates two types of beams coupled to an optical fiber patterned with well-designed ring-like gratings. Here, we have developed the power-in-bucket model to analyze the coupling efficiencies $\eta$ of structure-enhanced fibers and bare ones under large-angle multiple illumination conditions. Guided by the PIB model, $\eta$ could increase from 0.332022 to 0.5102 under the optimal coupling scenarios. The proposed model offers valuable guidance in designing highly efficient light-coupling systems using a beam-nanostructure-fiber configuration.
\end{tocentry}

\begin{abstract}
Well-designed nanostructures on fiber facets can boost wide-angle light coupling and thus gain considerable attention because of the potential for intensive applications. However, previous theories commonly concentrate on the configurations of the bare waveguide, lacking full consideration of structure-assisted couplings. Here, a power-in-bucket (PIB) model is introduced to explore the coupling behavior of structure-modified waveguides. The analytical model investigates two representative coupling scenarios, including Gaussian beam and plane wave excitation. The PIB-computed coefficient $\eta$ enhancements agree well with the experimental values, especially for the multiple-mode fibers under large-angle illuminations. Using PIB to optimize the beam-fiber parameters, we show that at the incidence angle of 37$^\circ$, $\eta$ could increase from 0.3320 to 0.5102 by the identical ring gratings. Overall, the proposed model provides a useful account of the mechanism of grating-aided light couplings. These findings would be of great help in designing structure-enhanced probes for wide-angle broadband light collection applications.  
\end{abstract}

\section{Introduction}

In a digitalized society, fiber networks are the cornerstone for supporting large-scale data transportation~\cite{snyder2012optical,saleh1991fundamentals}. One of the key functionalities behind the networks lies in fiber-based interconnections, where highly efficient fiber input and output are always demanded. Being a common type of optical input, the free light coupled to the fiber is a major technical indicator for high-performance optical systems. So far, numerous demonstrations alongside theoretical calculations have been carried out to analyze fiber-based coupling behaviors~\cite{jones1965coupling,saruwatari1979semiconductor,niu2007coupling,son2018high,gu2022investigations}. \par

While the fiber waveguide exhibits underachievement in coupling efficiency $\eta$, especially for wide-angle incidences. For instance, the $\eta$ of single mode fiber 28 (SMF-28) is much lower than $10^{-5}$ once the incidence angle $\theta$ exceeds 25$^{\circ}$~\cite{wang2019boosting}. Such a limited capacity hinders fiber usage in extensive applications like scanning near-field microscopy~\cite{xie2023large}, wide-view endoscopy~\cite{choi2012scanner}, Raman spectroscopy~\cite {latka2013fiber}, and optical connectivity~\cite{opticalIO,yermakov2025fiber}.\par  

To enhance the waveguide performance, recent years have seen emerging interest in the concept of structure-enhanced fibers~\cite{wang2018nanotrimer,wang2019boosting,xiong2020multifunctional,li2023metafiber}. By patterning well-engineered structures on fiber facets, this non-invasive method significantly upgrades their properties and thus broadens the applied scenarios. The representative works include but are not limited to the metalens-integrated fiber tip for optical trapping~\cite{plidschun2021ultrahigh} and spring-tipped waveguide probes for force sensing~\cite{shang2024fiber}.\par

Regarding wide-angle light collections, various kinds of micro-shapes have been implemented on the end faces of fibers. To date, the effective collecting angle is expanded up to 85$^{\circ}$ using the gold dot array~\cite{wang2019boosting,wang2021nanograting}. And the percentage level of $\eta$ is realized by 3D-printed polymer gratings~\cite{yermakov2020nanostructure,yermakov2023advanced}. Moreover, multiple wavelengths covering the single-mode and multi-mode ranges are achieved with the help of algorithm-conceived asymmetrical gratings~\cite{wang2024genetic,wang2024fiber}. All the progress solidly paves the way toward a probe capable of broad-bandwidth, wide-angle, and high-efficiency light gathering.\par

However, there is a lack of analytical or semi-analytical models that can analyze and guide the design of such structure-assisted probes. Previous theories focus on simplified coupling occasions, where the configurations of the bare waveguide alongside normal incidences are mainly demonstrated~\cite{saruwatari1979semiconductor}. Also, regarding fiber multi-mode and high-order beam excitations, the commonly-used efficiency analysis methods (such as coupled-mode theory) require summing up the values of every mode~\cite {niu2007coupling}. Additional numerical calculations must be conducted on a case-by-case basis, resulting in tedious and substantial workloads. Hence, developing general and easy-to-use models would be of help in understanding the mechanism of grating-aided light couplings and promoting light-collection fiber designs.\par

To address the above issue, we propose a power-in-bucket (PIB) model to inspect and enhance waveguide-based light-coupling performance. Originally, PIB is considered a key value for evaluating beam quality~\cite{pib2024edmund}. Here this method is extended as an intensity-based numerical model to describe the light proportion coupled to the waveguide. To demonstrate its general applicability, the nanostructure-integrated waveguides and a bare one under varied illumination conditions (including Gaussian beam and plane waves) are discussed, where both single-mode and multimode regimes are covered. Compared to the experimental data, the PIB-projected lineshapes display a great degree of similarity, while the rigorous theory fails to reproduce the measurement. In particular, with the help of ring-like gratings, the efficiency enhancement at wide incidence angles and multiple wavelengths (i.e., 650 and 1550 nm) is well-matched by the PIB model. At last, we employ the PIB model to refine beam-to-fiber coupling arrangements, in which grating-enabled efficiency $\eta$ is further promoted from 0.3320 to 0.5102. With the developed PIB model, one could optimize structure-functionalized waveguides so as to construct high-end low-loss optical systems as well as networks. Further proper modifications to the PIB model can promote broad emerging fields such as fiber-chip intercommunication \cite{wang2024precise} and on-chip couplings \cite{im2025chip}. \par

\section{Results and discussion}

\subsection{The PIB model for Gaussian beam couplings}
\subsubsection{Gaussian beam couplings based on bare waveguides}
\begin{figure}[ht!]
\centering\includegraphics[scale = 0.5]{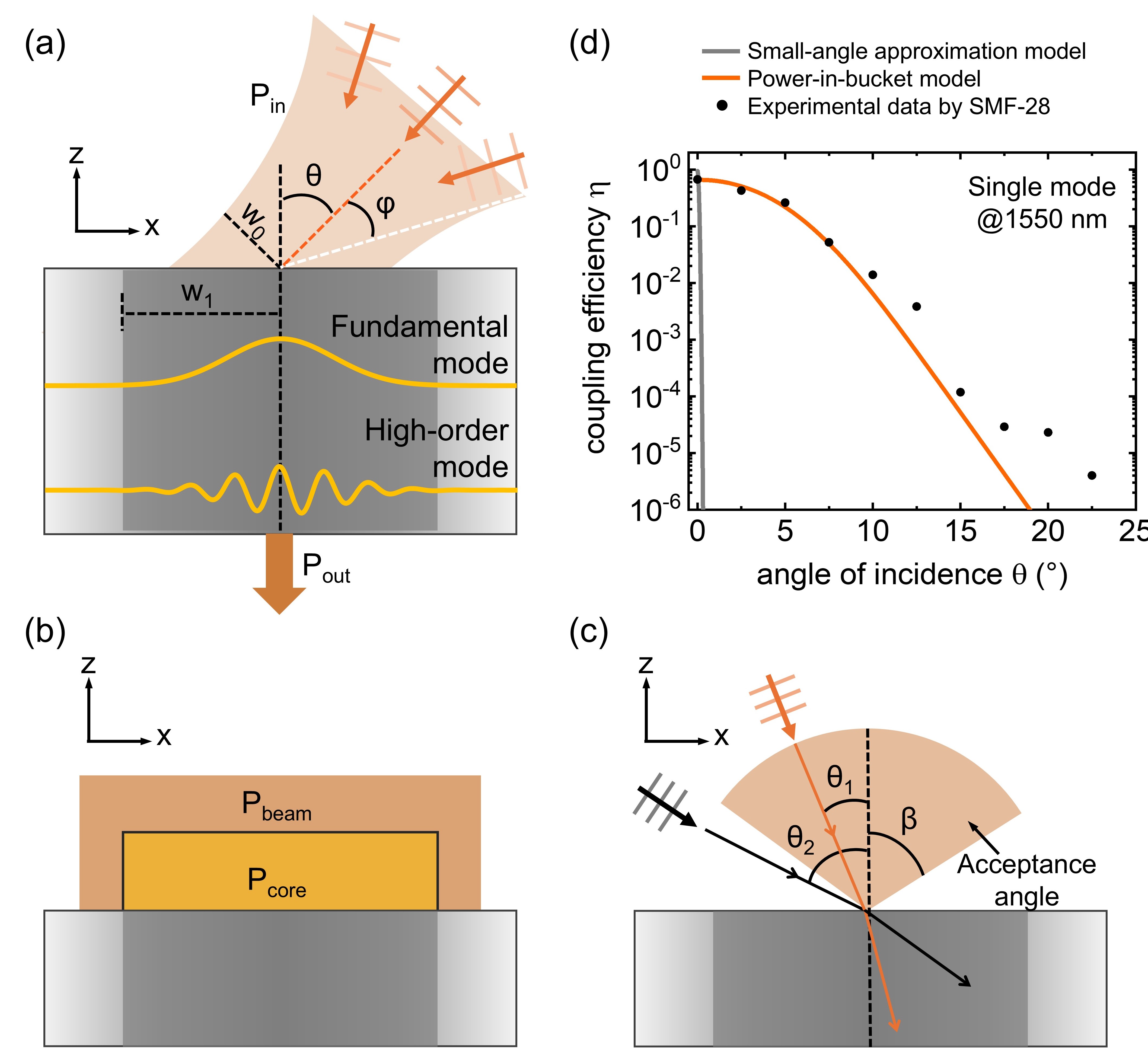}
\caption{The power-in-bucket (PIB) model for waveguide coupling behavior analysis. (a) A sketch illustrating that a focused Gaussian beam is incident on the end face of an optical fiber at an angle of $\theta$. (b-c) Two sketches correspond to PIB-related parameters of power ratios of $T_{beam}$ and $T_{NA}$, respectively. (d) A diagram comparing the coupling efficiency using the PIB model (orange line) and the rigorous small-angle-approximated model (light gray line), respectively. The black dots indicate the measured data of the bare single-mode fiber 28 (SMF-28).}
\label{fig-pib}
\end{figure}

PIB is a typical metric to assess the quality of laser beams~\cite{pib2024edmund}.In general, PIB denotes how much power is integrated over a 'bucket', where the bucket is often referred to as a specific spot. In terms of PIB for coupling investigations, the integrated power can be understood as the light amount entering the waveguide core, and the 'bucket' is designated as the overall incoming power~\cite{niu2007coupling}. Here the PIB model is further introduced in Fig. \ref{fig-pib}. A focused Gaussian beam (incoming power: $P_{in}$) impinges on the waveguide (mode radius: $w_1$), and partial light could be collected (transmitted power: $P_{out}$). The coupling efficiency $\eta (\theta)$  defined as $P_{out}/P_{in}$ is a function of the incident angle $\theta$, given the Gaussian beam parameters  the wavelength of $\lambda$, the beam half-width of $w_0$ and the divergence angle of $\psi$. \par

The PIB approach essentially accounts for two fundamental limitations in fiber coupling efficiency calculations, with one being the geometric constraint between the beam and the core, and another being the angular constraint of the fiber's numerical aperture (NA). Correspondingly, two preliminary coupling  parameters of $T_{beam}$ and $T_{NA}$ need to be defined prior to determining the overall coefficients $\eta_{pib}$. As can be seen from Fig.\ref{fig-pib} (b), the first one $T_{beam}$ is the ratio of the energy reaching the fiber’s core (namely $P_{core}$) over the total power reaching the fiber endface ($P_{beam}$), i.e., $T_{beam} = P_{core}/P_{beam}$. The complete form of the Gaussian $T_{beam}$ can be expressed as follows.

\begin{equation}
    T_{\text {beam}}(\theta)=\frac{\int_{-w_{1}}^{w_{1}} \frac{w_{0}^{2}}{w(x, \theta)^{2}} \exp \left(\frac{-2(x \cos (\theta))^{2}}{w(x, \theta)^{2}}\right) d x}{\int_{-\infty}^{\infty} \exp \left(-2\left(x / w_{0}\right)^{2}\right) d x}
    \label{tbeam}
\end{equation}

The second parameter $T_{NA}$ is explained as the model coupling between the Gaussian beam and the waveguide mode, as illustrated in Fig.\ref{fig-pib} (c). Here, the Gaussian beams are treated as a superposition of plane waves with different propagation angles. Plane waves with small angles (e.g., $\theta_1$) result in a high proportion of guided transmission (see the orange arrow). Conversely, plane waves with larger angles (e.g., $\theta_2$) lead to greater energy loss (see the black arrow), since the majority of light is beyond the acceptance angle $\beta$ and thus fails to be coupled into waveguides. Therefore, $T_{NA}$ can be quantified as the subsequent equation.

\begin{equation}
    T_{NA}(\theta)=\frac{\int_{-\pi / 2}^{\pi / 2} \eta(i) \exp \left(\frac{-2(i-\theta)^{2}}{\varphi^{2}}\right) d i}{\int_{-\pi / 2}^{\pi / 2} \exp \left(\frac{-2 i^{2}}{\varphi^{2}}\right) d i}
    \label{tbeta}
\end{equation}

where $\exp \left(\frac{-2(i-\theta)^{2}}{\varphi^{2}}\right)$ describes the intensity distribution of the plane waves (decomposed from the Gaussian beam) at different incident angles of $i$. $\eta(i)$ is an empirical formula, representing the energy proportion of transmission modes excited by the plane wave. A detailed discussion of $\eta(i)$ is provided in the Supplementary Information (SI).

Therefore, $\eta_{pib}$ is written as:

\begin{equation}
\eta_{pib}=T_{beam}\cdot T_{NA}
\label{eq-pib}
\end{equation}

Fig. \ref{fig-pib} (d) presents PIB-calculated coupling efficiencies $\eta_{pib}$ (orange line) with experimental data (black dots). A bare SMF-28 ($w_1 = 5.05~\mu m$) is illuminated by a focused Gaussian beam ($\lambda$ = 1.55 $\mu m$, $w_0 = 7.05~\mu m$) supposing that there is no beam-fiber misalignment. For comparison, a rigorous model using the small-angle approximation method is also plotted (see the reference \cite{jones1965coupling} and SI for the detailed model). In contrast to the rigorous $\eta$ (light gray line), $\eta_{pib}$ is much more consistent with the measurements, which confirms the correctness and accuracy of our proposed methods.\par

\subsubsection{Gaussian beam couplings by structured waveguides}
\begin{figure}[ht!]
\centering\includegraphics[scale = 0.5]{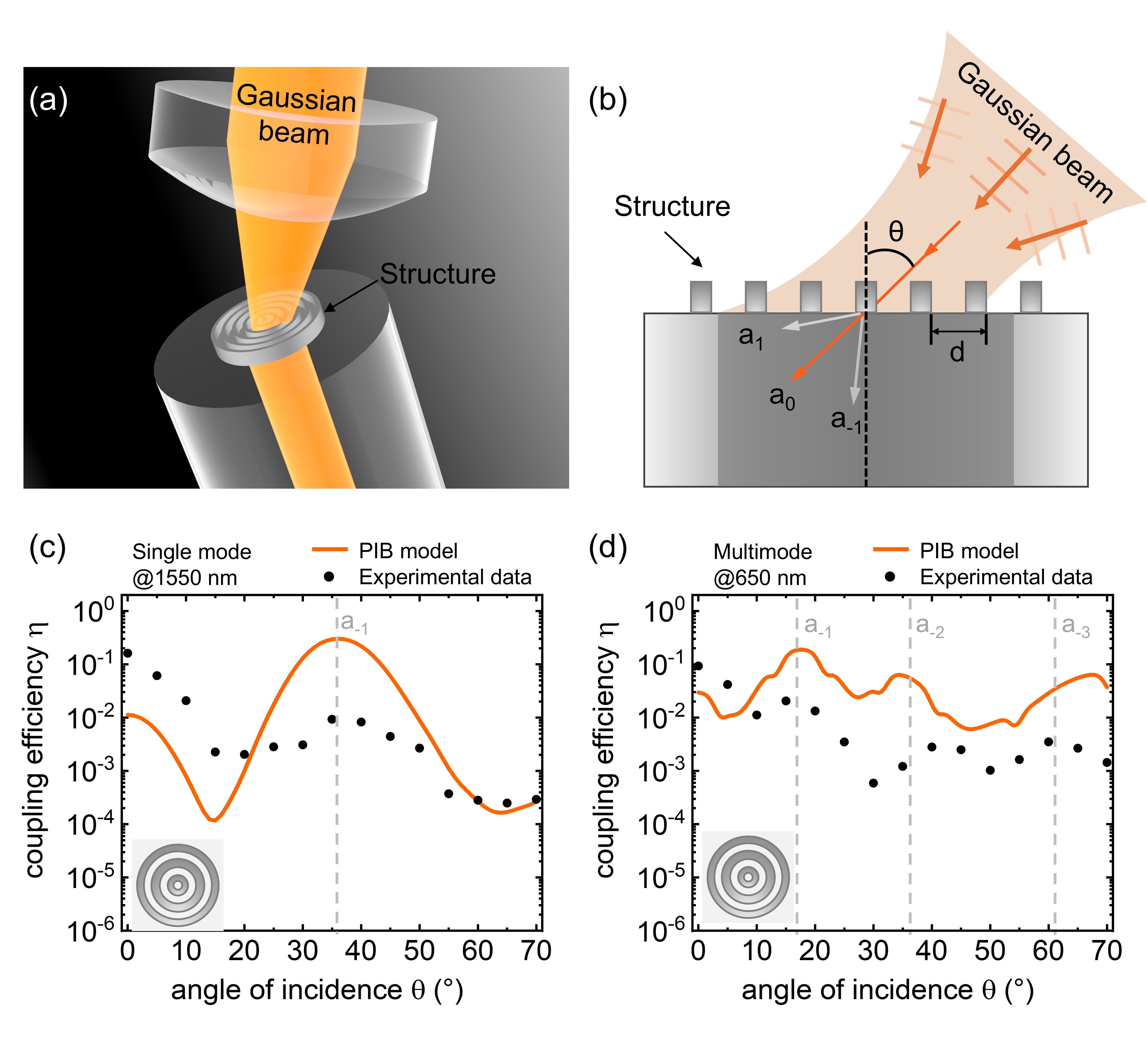}
\caption{Boosting Gaussian beam-illuminated coupling efficiency using structure-enhanced waveguides. (a) The diagram of a ring-like grating patterned on the fiber facet. (b) A 2D model showing that the gratings split incoming Gaussian light into several diffraction orders, and the additional light can be coupled into the waveguide. (c-d) Wide-angle coupling coefficient enhancement using the polymer ring grating at a wavelength of 1550 nm and 650 nm, respectively. The black dots represent the measured efficiencies using grating-integrated SMF-28.}
\label{fig-gaussian}
\end{figure}

To lift the coupling efficiency on demand, one promising solution is to fabricate well-crafted nanostructures on the fiber endface. The operation principle is plotted in Fig. \ref{fig-gaussian} (a) and (b), where a Gaussian beam approaches the grating-patterned fiber tip at an angle of $\theta$. The periodical structure (pitch of $d$) divides the beam into several diffraction orders (here diffraction efficiencies are $a_{-1}$, $a_0$, and $a_{+1}$, respectively). In this case, by combining with the grating equation, the parameter $T_{NA}$ can be modified as $T_{NA_{-}\text {grating}}$. 

\begin{equation}
    T_{NA_{-} \text {grating }}(\theta)=\frac{\sum_{l} \int_{i_{l_{-}lower} }^{i_{l_{-}upper}} \exp \left(\frac{-2(i-\theta)^{2}}{\varphi^{2}}\right) \sum_{m} a_{m}(i) \eta_{m}(i) d i}{\int_{-\pi / 2}^{\pi / 2} \exp \left(\frac{-2 i^{2}}{\varphi^{2}}\right) d i}
\end{equation}

The term of $ \exp \left(\frac{-2(i-\theta)^{2}}{\varphi^{2}}\right)$ is related to the plane waves. ${i_{l_{-}lower}}$ and ${i_{l_{-}upper}}$ is the lower and upper limits of the $l^{th}$ integration segment of corresponding plane waves. The section of $a_{m}(i) \eta_{m}(i)$ is determined by the grating effect. $m$ denotes the diffraction order and $a_{m}(i)$ represents the diffraction efficiency of the $m^{th}$ order. $\eta_{m}(i)$ corresponds to the coupling efficiency of the $m^{th}$ order diffracted wave. More detailed illustrations on ${i_{l_{-}lower}}$, ${i_{l_{-}upper}}$, $a_{m}(i)$, and $\eta_{m}(i)$ can be examined in SI.\par

Therefore, the PIB theory of grating-assisted efficiency $\eta_{grating}$ can be described in the following form.

\begin{equation}
    \eta_{grating}(\theta)=T_{beam} \cdot T_{NA_{-} \text {grating }}
    \label{pib-grating}
\end{equation}

To verify the structure-modified PIB model, an SMF-28 (NA$_{fiber}$: 0.1084, core radius $w_1$: 5.055 \textmu m) integrated with polymer-made ring-like gratings is characterized via the focused Gaussian beam illuminations (more information on the structure and experimental setup can be seen in SI). Fig.\ref{fig-gaussian} (c-d) presents PIB-computed coupling efficiency (orangle line) together with the measured counterparts (black dots). Note that both single-mode (1550~nm) and multi-mode range (650~nm) of SMF-28 are taken into account, where the beam waist radius $w_0$ is estimated as 7.05 \textmu m  and 6.35 \textmu m, respectively. \par

It is noted that coupling coefficients acquired within both fiber single-mode and multi-mode ranges can be well-fitted by the PIB model. The black dots in two graphs below display varied lineshapes due to the wavelength-dependent grating modifications. In addition to the value maximum at normal incidence ($\theta$ = 0°), only one enhanced peak (around 35°) is observed for $\lambda$ = 1550 nm, while three less-distinguished value improvements (around $\theta$ = 20°, $\theta$ = 40°, and $\theta$ = 60°) emerge in the case of $\lambda$ = 650 nm. Such distributions are indeed captured by the PIB theory, where both -1$^{st}$, -2$^{nd}$, and -3$^{rd}$ diffraction orders contribute at shorter wavelengths, while only the -1$^{st}$ order is relevant when the pitch value approximates the operation wavelength. Besides, the enhanced angular intervals can be simply derived from the grating equation, where the vertical lines, along with labels of $a_{-1}$, $a_{-2}$, and $a_{-3}$, indicate the corresponding diffraction orders.\par

\subsection{The PIB model for plane waves couplings}
\subsubsection{Plane wave couplings of bare waveguides}
\begin{figure}[ht!]
\centering\includegraphics[scale = 0.5]{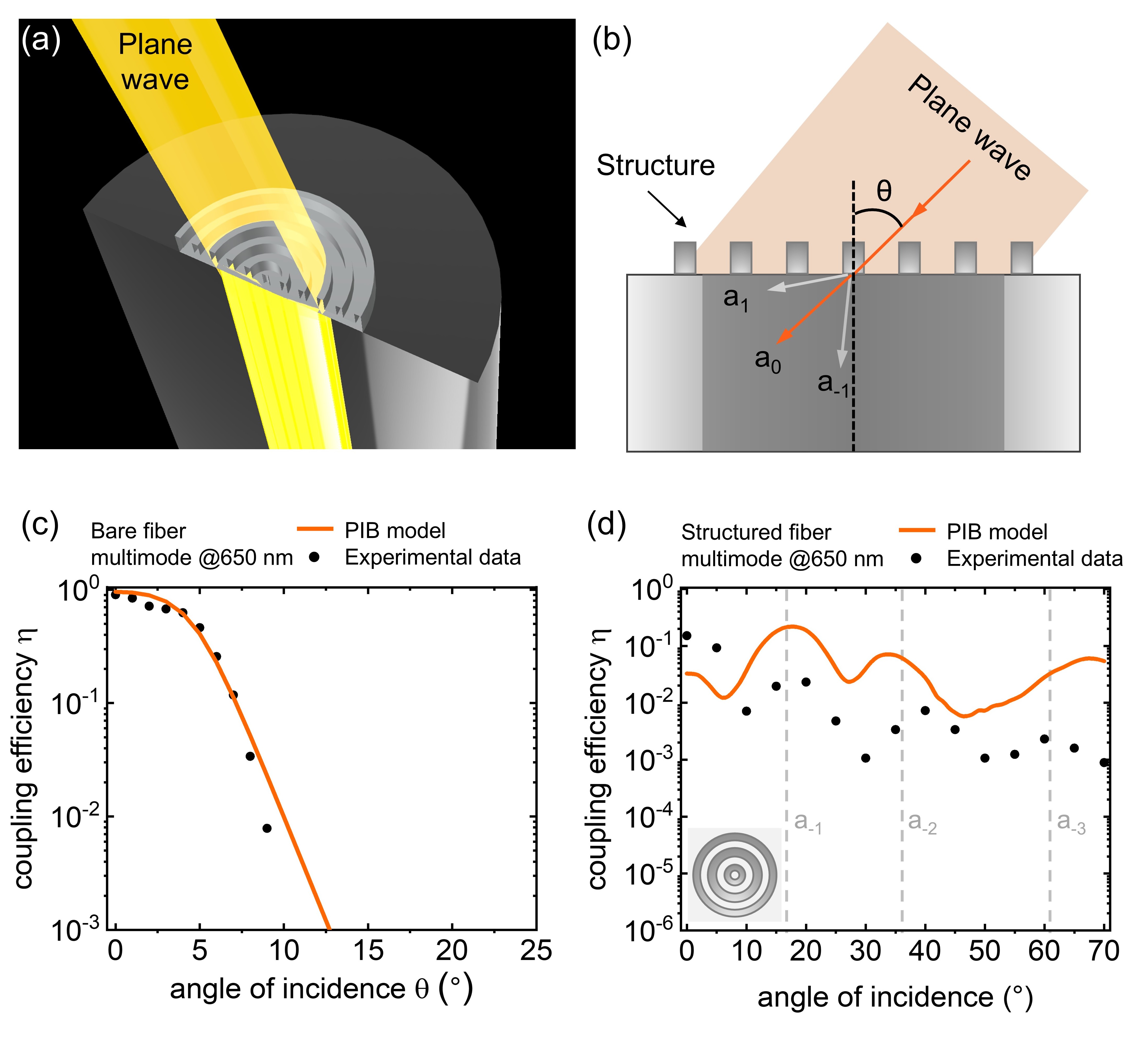}
\caption{Coupling efficiencies improvement of the grating-fiber under plane wave excitation. (a-b) A sketch along with the 2D model mimicking a plane wave incident on the grating-printed fiber endface. The coupling efficiency comparison between a bare SMF-28 fiber (c) and the grating-enabled one (d). The orange line and black dots originate from the PIB calculations and measurements, respectively.}
\label{fig-plane}
\end{figure}

Since the PIB approach is an intensity-based model, in principle, any incoming light that can be approximated by the superpositions of plane waves is able to be inspected by the theory. Hence, to examine its broad applicability, here we further perform the PIB calculation under the plane wave excitation. Fig. \ref{fig-plane} plots the PIB analysis for a fiber waveguide excited by a plane wave. We first check the bare fiber coupling case, and the grating-reinforced fiber couplings would be explained later. \par

As discussed in eq.\ref{eq-pib}, two main components of $T_{beam}$ and $T_{NA}$ need to be modified as illuminations shift from a Gaussian beam to a plane wave. $T_{beam}$ in eq.\ref{eq-pib} corresponds to the power ratio of the energy reaching the fiber’s core (area of $A_{core}$) over the total beam size (area of $A_{beam}$). Hence, here $T_{beam}$ becomes a size ratio bewteen $A_{core}$ and plane wave beam size $A_{plane_{-}wave}$, i.e.,
\begin{equation}
    T_{beam}=A_{core}/A_{plane_{-}wave}
    \label{pib-plane-Tbeam}
\end{equation}

For convenience, $T_{beam} = 1$ is applied throughout the article, considering the experimental configuration in Fig.\ref{fig-plane}.  \par
As for $T_{NA}$, the original beam divergence angle of $\psi$ in Equation \ref{tbeta} becomes extremely small, where the term of $
    \frac{\exp \left(\frac{-2(i-\theta)^{2}}{\varphi^{2}}\right)}{\int_{-\pi / 2}^{\pi / 2} \exp \left(\frac{-2 i^{2}}{\varphi^{2}}\right) d i}
 $ can be approximated by a delta function. Therefore, connecting with Eq.\ref{pib-plane-Tbeam}, the plane wave coupling efficiency $\eta_{\text{planewave}}$ can be represented as: 
 
\begin{equation}
\eta_{\text{planewave}}(\theta)=T_{beam}\cdot T_{NA}=\Bigl\{\frac{-1}{1+\exp (-(1/V)^{1/10} (\theta-\psi))}+1 \Bigl\}\Bigl\{\frac{1}{1+\exp (-(1/V)^{1/10}(\theta+\psi))}\Bigl\}
\label{pib-planewave-barefiber}
\end{equation}

where the V  is calculated using the formula $2\pi w_{1}NA/\lambda$, in which $w_{1}$ represents the core radius of the fiber and $\lambda$ denotes the wavelength of the incident light. $\psi$ is the numerical angle associated term, and more detailed discussions can be found in the SI.  \par

\subsubsection{Plane wave couplings of grating-enhanced waveguides}

The interaction between a plane wave and a grating-patterned fiber becomes the focus in this section.  As illustrated in Figure \ref{fig-plane} (b), when a plane wave is incident on the fiber end-face, it is diffracted into multiple waves propagating at certain directions. The total coupling efficiency $\eta_{\text{grating}_{-}\text{{planewave}}}$ is obtained by:

\begin{equation}
    \eta_{\text{grating}_{-}\text{{planewave}}}(\theta)= T_{beam}\cdot \sum_{m} a_{m} \eta_{m}\left(i_{\text {out }}\right)
    \label{pib-plane-grating}
\end{equation}

where $a_{m}$ denotes the diffraction efficiency of the m$^{th}$ order beam, while $\eta_{m}$ corresponds to the coupling efficiency of the m$^{th}$ order diffracted plane wave. $i_{out}$ represents the output angle of the $m^{th}$ diffracted order, that subsequently serves as the input angle for the fiber couplings.\par

Eq.\ref{pib-planewave-barefiber} and Eq.\ref{pib-plane-grating} are experimentally studied in Fig.\ref{fig-plane} (c-d), where two plots correspond to the coupling efficiency ($\lambda =650 nm $) based on a bare SMF and grating-patterned one, respectively. In Fig. \ref{fig-plane} (c), the orange curve computed by eq.\ref{pib-planewave-barefiber} is in line with measured data, which generally decays as the input angle increases. While the coefficients are substantially lifted in Fig. \ref{fig-plane} (d) due to the presence of well-designed gratings. Three amplitude improvements at $\theta$ of about 15°, 35°, and 60° can be noted (see the vertical dashed lines), which are reproduced by the eq. \ref{pib-plane-grating}. These results align well with the proposed theoretical predictions, extending the applicability of the PIB model to more incidence situations.\par

\subsection{PIB-optimized configurations elevating coupling coefficients}

\begin{figure}[ht!]
\centering\includegraphics[scale = 0.5]{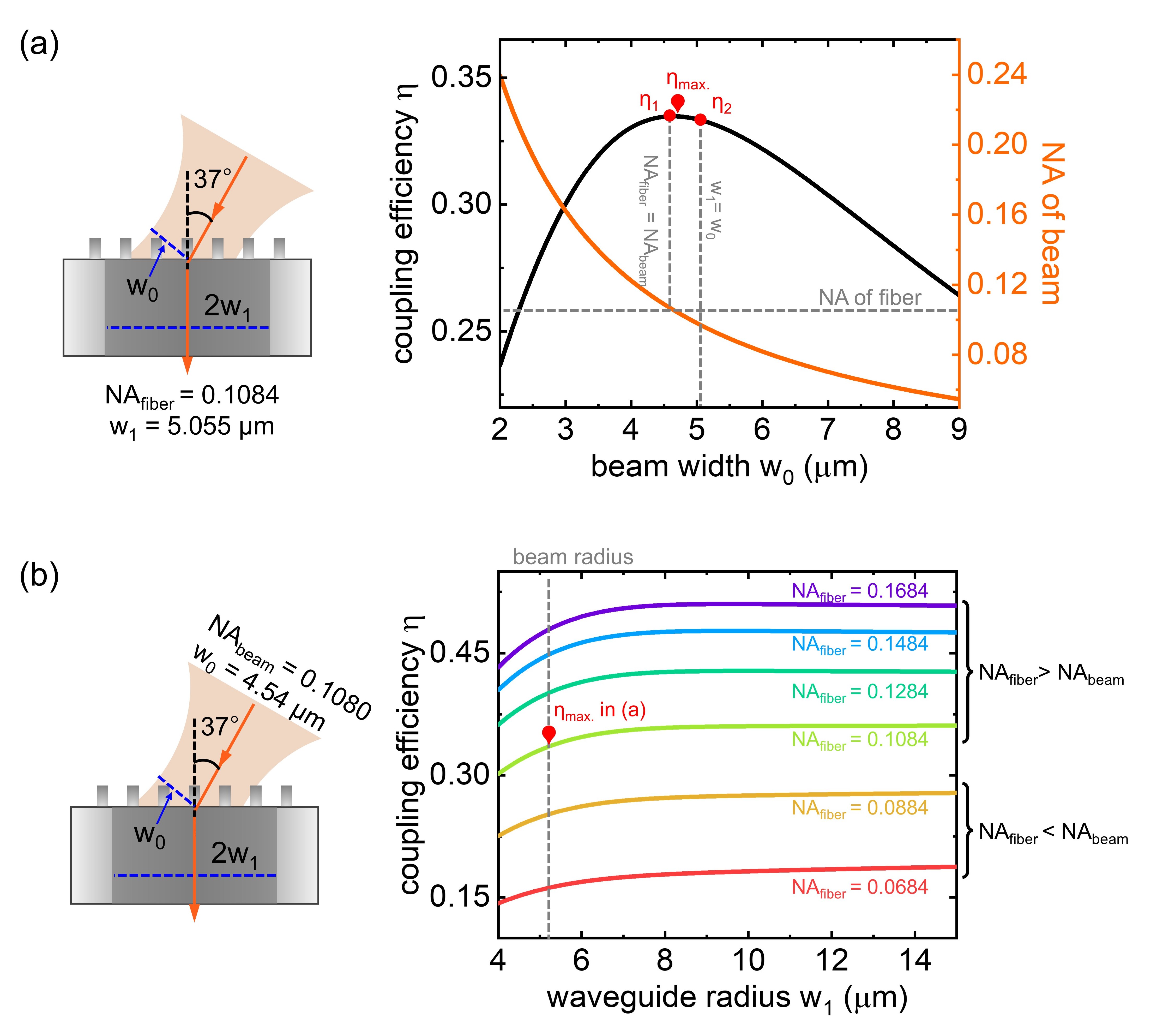}
\caption{Further improving coupling efficiency $\eta$ by modifying beam and waveguide property. The grating (same as Fig.~\ref{fig-gaussian}) is excited under $\theta = 37^{\circ}$ incidence light ($\lambda$  = 1550 nm) for both graph. (a) $\eta$-vs-$w_0$. The NA of fiber is fixed at 0.1084 with $w_1 = 5.055~\mu m$, while $NA_{beam}$ and half-width $w_0$ are tuned. (b) $\eta$-vs-$w_1$. The beam is arranged as $NA_{beam}$ = 0.1080 and $w_0=4.54 \mu m$. Six types of $NA_{fiber}$ along with increased radii $w_1$ are selected to examine their influence on $\eta$.}
\label{fig4}
\end{figure}

PIB model can guide the construction of a high $\eta$ system. In addition to the nanostructure modification, the beam and fiber properties could impact the $\eta$ as well. For demonstration purposes, Fig. \ref{fig4} explores coupling coefficient variations using the identical grating and illumination conditions as Fig. \ref{fig-gaussian} (c). Note that, Fig. \ref{fig4} (a) fixes the fiber as $NA_{fiber}$ = 0.1084 and $w_1$ = 5.055 $\mu$m, whereas Fig. \ref{fig4} (b) reverses the situation. The beam parameters are constants ($NA_{beam}$ = 0.1080 and $w_0$ = 4.54 $\mu$m) while the fiber's parameters (including $NA_{fiber}$ and core radius $w_1$) changes accordingly.\par

Fig. \ref{fig4} (a) highlights that despite the waveguide's small NA, the large coupling coefficients at the wide input angle can be obtained with the assistance of ring gratings. In detail, as the beam half-width $w_0$ increases, $\eta$ first rises to the maximum value of $\eta_{max}$ and then slowly declines (see the navy curves). Located near the $\eta$ peak scope, $\eta_1$ and $\eta_2$ denote the coupling coefficients gained under the conditions of $NA_{fiber}=NA_{beam}$ and $w_1=w_0$, respectively. Note that the optimal efficiency of $\eta_{max}$ is near two cases. In practical terms, the NA-matched principle can be employed to build a high-coupling low-loss system, since the value gaps among $\eta_1=0.3320$, $\eta_2=0.3297$, and $\eta_{max}=0.3321$ are much limited.\par

Adopting the arrangement of $\eta_{max}$ in Fig. \ref{fig4} (a), Fig.~\ref{fig4} (b) continues to pursue the larger $\eta$ by improving fiber NA and radius $w_1$. Classified as two types of NA combinations (e.g., $NA_{fiber}>NA_{beam}$ and $NA_{fiber}<NA_{beam}$), all efficiency curves increase as a function of waveguide radius $w_1$, where the first one with $NA_{fiber} = 0.1684$ tops all (see the purple line). It is noteworthy that neither $NA_{fiber}=NA_{beam}$ nor $w_1=w_0$ yields the highest $\eta$. In fact, the combination of the maximum values of $NA_{fiber}=0.1684$ and $w_1=9.6 \mu m$ produces an optimal $\eta$ of 0.5102. Hence, the combination of higher $NA_{fiber}$ and $w_1$ is suggested to create the best-fit coupling condition. \par

Here, we would like to mention that the current PIB model is a 2D semi-analytical model. Additional modulation factors, such as the lensing effect caused by ring gratings, are beyond the current discussion. But the PIB can be employed to quickly compute coupling efficiency to meet various usage needs. Further combined with intelligent algorithms \cite{wang2021intelligent}, the PIB model is capable of constructing microstructures for real applications like Raman spectroscopy \cite{gu2025}.\par

\section{Conclusion}

The research set out to establish a theoretical model to devise the structure-patterned waveguides capable of collecting broadband, wide-angle incidence lights. For this purpose, utilizing two power ratios to replicate the aforementioned coupling scheme, the analytical PIB model is proposed, where various coupling configurations by changing, e.g., the beam type and wavelength, are examined in detail. Particularly, a representative structure of ring gratings is exemplified to lift efficiency $\eta$, where the $\eta$ enhancements under both Gaussian and plan waves are all reproduced by the PIB model. We further illustrate the potential to increase coupling coefficients employing the PIB-adjusted beam-fiber parameters. Under the combination between $NA_{fiber}>NA_{beam}$ and $w_{fiber}$$>$$w_{beam}$, the optimal $\eta$ of 0.5102 at $\theta = 37 ^\circ$ can be achieved by the ring-grating-empowered waveguides, where the original counterpart is only improved to 0.3320. In conclusion, the proposed model offers valuable guidance for the design and optimization of coupling-used structures, enabling wide-angle broadband light gathering and inter-connections for emerging fields.

\section{ASSOCIATED CONTENT}

\subsection{Supplementary information}
The Supporting Information is available free of charge at XXXX. 

\subsection{Data availability statements}

The data that supports the findings of this study are available within the article [and its supplementary material].

\subsection{Author Contributions}
N.W. conceived the idea, and W.L. developed the model. W.L. and N.W. performed coupling analysis and data comparisons. W.L., Y.G., and J.W. obtained the experimental data. All authors contributed to the discussions on data analysis and manuscript writing. 
\subsection{Funding}
The work was supported by the HIAS project (No. 2024HIAS-Y009), the National Key Research and Development Program of China (Grant No. 2022YFA1405200), the National Natural Science Foundation of China (NSFC Grants Nos. 12474393 and 12174340), and the National Key Laboratory of Spatial Information Laser Foundation (No. LSI2025WDZC12).
\subsection{Competing interests}
The authors declare no competing interests.
\subsection{}

\bibliography{achemso-demo}
\end{document}